Preprint

# Multi-faceted Methodology for Coastal Vegetation Drag Coefficient Calibration: Implications for Wave Height Attenuation

Erfan Amini[1], Reza Marsooli[1*], Mehdi Neshat[2]

*1. Department of Civil, Environmental, and Ocean Engineering, Stevens Institute of Technology, Hoboken, New Jersey, USA*
*2. Faculty of Engineering and Information Technology, University of Technology Sydney, Ultimo, NSW, Australia*

*Abstract*

The accurate prediction of wave height attenuation due to vegetation is crucial for designing effective and efficient natural and nature-based solutions for flood mitigation, shoreline protection, and coastal ecosystem preservation. Central to these predictions is the estimation of the vegetation drag coefficient. The present study undertakes a comprehensive evaluation of three distinct methodologies for estimating the drag coefficient: traditional manual calibration, calibration using a novel application of state-of-the-art metaheuristic optimization algorithms, and the integration of an established empirical bulk drag coefficient formula (Tanino and Nepf, 2008) into the XBeach non-hydrostatic wave model. These methodologies were tested using a series of existing laboratory experiments involving nearshore vegetation on a sloping beach. A key innovation of the study is the first application of metaheuristic optimization algorithms for calibrating the drag coefficient, which enables efficient automated searches to identify optimal values aligning with measurements. We found that the optimization algorithms rapidly converge to precise drag coefficients, enhancing accuracy and overcoming limitations in manual calibration which can be laborious and inconsistent. While the integrated empirical formula also demonstrates reasonable performance, the optimization approach exemplifies the potential of computational techniques to transform traditional practices of model calibration. Comparing these strategies provides a framework to determine the most effective methodology based on constraints in determining the vegetation drag coefficient.

---

* Corresponding author.
*E-mail address:* rmarsool@stevens.edu

## 1. Introduction

The pivotal role of coastal vegetation in attenuating waves is well established within the scientific literature (Lg et al., 2022; Moraes et al., 2022; Quataert et al., 2015; Z. Zhu et al., 2020). A precise understanding of wave-vegetation interactions and their resultant impact on wave energy is of paramount importance for the design of vegetated buffers against waves. The attenuation of waves in the presence of nearshore vegetation is dependent on both the physical and biological properties of the vegetation, such as the species, the density and flexibility of the vegetation, and the submergence depth (Liu et al., 2023; Miller et al., 2022; Rosenberger & Marsooli, 2022; W. M. Wu et al., 2011). Each of these factors contributes uniquely to the vegetation drag coefficient, a measure of the resistance a wave encounters while propagating through a vegetated area. In addition to the characteristics of the vegetation itself, bathymetry, and the features of the incident waves, such as height and period, affect the drag exerted by the vegetation and thus the subsequent wave attenuation (Marsooli et al., 2016; Tanino & Nepf, 2008). The complexities of these relationships make wave height attenuation predictions a sophisticated task that necessitates careful estimation of drag coefficients for optimal outcomes. Moreover, the ability to accurately predict wave height attenuation is vital in the development and application of effective coastal management strategies (Lee et al., 2021). These predictions inform the design and implementation of nature-based shoreline protection measures and provide valuable input for coastal zone planning and management (Clark, 1997). This is particularly important given the increase in extreme weather events linked to climate change, which pose a significant risk to coastal communities and ecosystems (Amini, Marsooli, et al., 2023; Jamous et al., 2023; Marsooli & Lin, 2020; Nicholls et al., 2018).

Vegetation drag coefficient ($C_d$) provides a measure of the force exerted by vegetation on a wave and is central to the predictions of wave height attenuation. The drag coefficient is influenced by both the physical characteristics of the vegetation and the hydrodynamic conditions, making its calibration a complex task (Familkhalili & Tahvildari, 2022; Liu et al., 2023). Traditionally, manual calibration has been used to determine the drag coefficient. However, this method is labor-intensive and time-consuming, and sometimes lacks the precision required for accurate wave height attenuation predictions (Anderson & Smith, 2015; Gijón Mancheño et al., 2021). This method also requires expertise on the part of the numerical modeler. Consequently, the results obtained from this approach could vary widely based upon the environmental conditions and wave characteristics (Burger, 2005). A significant advance in this area came with the development of empirical formulations for bulk drag coefficients, such as the formulation proposed by (Mendez & Losada, 2004; Ozeren et al., 2014; Tanino & Nepf, 2008; L. Zhu et al., 2023), which estimate the drag coefficient as a function of hydrodynamic conditions and vegetation properties.

Many empirical formulas have been developed to estimate the bulk $C_d$ (Mendez & Losada, 2004; Ozeren et al., 2014; Tanino & Nepf, 2008; L. Zhu et al., 2023). Most of the formulas are developed based on the water particle velocities, which emerge in the Keulegan–Carpenter number (KC) or Reynolds (Re) number. For example, (Tanino & Nepf, 2008), "T&N" hereafter, conducted a comprehensive set of laboratory tests analyzing drag on cylinder arrays representing emergent vegetation. They proposed a bulk $C_d$ formula as a function of the volumetric fraction of vegetation and the flow velocity through the vegetated area. Compared to prior bulk coefficients, this formulation reduces overestimation of $C_d$ and shows closer agreement with measured data (Tanino & Nepf, 2008). Similarly, (Ozeren et al., 2014) used laboratory experiments to quantify wave attenuation by vegetations with different types, densities, and heights under regular and irregular waves. They presented a bulk formula for the vegetation drag coefficient as a function of Keulegan-Carpenter number.

In addition to manual calibration and the use of empirical formulas, a few recent works have sought to improve drag coefficient calibration in similar applications by incorporating advanced optimization algorithms for semisubmersible platform (Böhm et al., 2020) and scenario-based selection of drag coefficient (Liu et al., 2023). These algorithms, which utilize computational techniques to optimize complex systems, offer potential for enhanced accuracy in drag coefficient calibration. Unlike the manual calibration of the drag coefficient, which necessitates experience from the numerical modeler, these optimization algorithms automate the search across possible ranges of drag coefficients. This automated process enables the selection of the optimal drag coefficient that ensures the highest degree of agreement between numerical results and the measured dataset. This distinction not only streamlines the calibration process, but also could enhance the accuracy of wave attenuation predictions, thereby offering potential for future applications in coastal engineering. However, this research avenue is yet to be fully developed by novel and accurate optimization algorithms. To fill this gap in the current research, the present study incorporates two recently developed metaheuristic algorithms to find the optimal values for the drag coefficient.

The present study investigates how an optimization approach for calibrating vegetation drag coefficient performs compared to approaches based on a manual calibration and an empirical formula. The study incorporates the empirical bulk drag coefficient formula of (Tanino & Nepf, 2008) into the XBeach non-hydrostatic (XBNH) model (Roelvink et al., 2009). This represents an improvement to the XBNH model, given that the drag coefficient in the model is a predefined temporally constant value. The integration of the bulk drag coefficient formula allows the model to determine the drag coefficient as a function of hydrodynamic conditions, which can vary in time and space. By comparing traditional manual calibration methods, metaheuristic optimization algorithms, and the integrated bulk drag coefficient formula, this research sheds light on the multi-faceted strategies for estimating the drag coefficient.

Following this introduction, section 2 focuses on the methods employed in our study, beginning with an overview of the calibration techniques used, before providing in-depth descriptions of the manual calibration, the metaheuristic optimization algorithms, and the bulk drag coefficient formula methods. In section 3, we present our results and engage in a discussion to interpret these findings and link them with existing knowledge in the field. This section also compares and contrasts the three methodologies explored. Finally, in section 4, we draw conclusions from our findings and discuss the implications of our research for the wider field of coastal engineering, highlighting the significant potential for improving wave height attenuation simulations through innovative drag coefficient calibration strategies.

## 2. Methodology

In this study, we focus on the vegetation effects on the wave height, by suggesting a multi-faceted approach for calibrating the drag coefficient. This approach incorporates three faceted methodologies based on numerical resources, expertise, and availability of the measured data (Figure 1). Manual calibration of the drag coefficient offers the advantage of the numerical modeler's prior experience. However, it can be time-consuming, labor-intensive, and prone to human error, leading to results that can vary in quality depending on the modeler's skill level (Djeddi et al., 2022). As another part of this study's method, metaheuristic optimization algorithms excel in efficiency and scalability, allowing for automated searches across extensive parameter spaces for an optimal or near-optimal solution. However, optimization algorithms can be computationally demanding, lack physical interpretability, and may require fine-tuning to avoid local minima, thereby making them less straightforward to apply in some instances (Amini, Nasiri, et al., 2023). Incorporating an empirical bulk drag coefficient formulation simplifies the calibration process and adds an empirically validated layer of reliability. However, the approach might be limited by the scope of the empirical formula itself, which may not be universally applicable across diverse hydrodynamic conditions or vegetation properties and therefore may result in less accurate estimations of the drag coefficient.

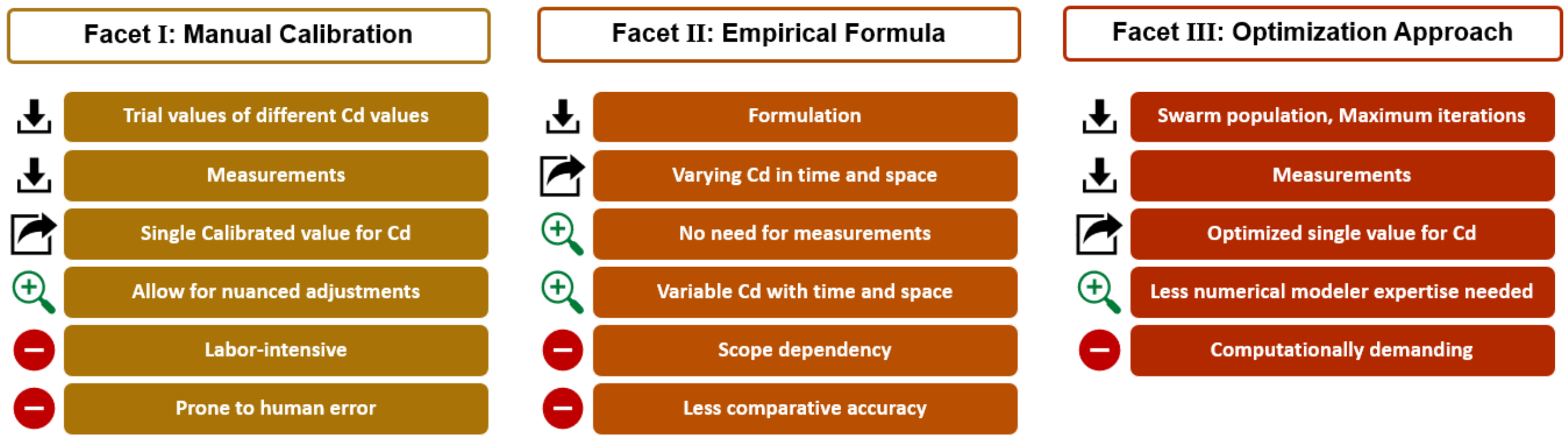

Figure 1. An overview of methodology facets and their input, output, advantageous (+), and drawbacks (-) for calibrating vegetation drag coefficient ($C_d$), all incorporated in the numerical simulations.

### 2.1. Numerical Model and Empirical Bulk Drag Coefficient

Numerical model XBeach Non-Hydrostatic (XBNH) (Roelvink et al., 2009) is adopted here to model the interactions between waves and coastal vegetation. The XBNH model is a depth-averaged phase-resolving wave model that solves nonlinear shallow water equations corrected for non-hydrostatic pressure effects. The vegetation is represented in the model as cylindrical elements. The associated vegetation-induced drag force is incorporated in the momentum equation as described by (van Rooijen et al., 2016; Yin et al., 2021). The wave energy is attenuated due to the exerted vegetation drag extracting energy from the wave-induced flow. The governing equations in XBNH are here simplified in one-dimensional shallow water equations for depth-averaged wave propagation (Equations 1 and 2). These equations represent the conservation of mass and momentum for wave propagation in the cross-shore direction.

$$\frac{\partial \eta}{\partial t} + \frac{\partial du}{\partial x} = 0 \tag{1}$$

$$\frac{\partial u}{\partial t} + u\frac{\partial u}{\partial x} = -\frac{1}{\rho}\frac{\partial(\overline{q} + \rho g \eta)}{\partial x} - \sqrt{\frac{gn^2}{d}}\frac{u|u|}{d} + \nu_h \frac{\partial^2 u}{\partial x^2} + \frac{F_{v,nh}}{\rho d} - \frac{\tau_{\beta,x}}{\rho d} \tag{2}$$

The spatial and temporal independent variables are denoted by $x$ and $t$. The water surface elevation is given by $\eta$, and $u$ represents the depth-averaged flow velocity. The local water depth is $d$. Horizontal viscosity is incorporated through the term $\nu_h$. The non-hydrostatic pressure contribution is captured by the depth-averaged dynamic pressure term $\overline{q}$, and $\tau_{\beta,x}$ is the bed shear stress. The vegetation exerts a drag force on the water flow, based on Equation (3). This drag force is proportional to the vegetation density, stem diameter, drag coefficient, and the flow velocity. The vegetation-induced hydrodynamic drag is represented by the source term $F_{v,nh}$, which integrates the vegetation drag over its height. As shown in prior works (Dalrymple et al., 1984; Ma et al., 2013; Marsooli & Wu, 2014; Wei & Jia, 2014), this vertically integrated vegetation drag can effectively model wave energy dissipation within the vegetation field.

$$F_v = \frac{1}{2}\rho C_D b_v N_v u_r |u_r| \tag{3}$$

In this context, $C_D$ represents the bulk drag coefficient. The term $b_v$ denotes the diameter of the vegetation stem, while $N_v$ signifies the density of vegetation, expressed as the count of stems per unit of area. The velocity $u_r$ is the relative velocity between the plant motion ($u_p$), and ambient water flow velocity ($u$), as it reads in Equation (4).

$$u_r = u - u_p \tag{4}$$

Generally, mangroves, dense forests, and idealized salt marshes can be considered rigid (Anderson & Smith, 2014), with $u_p$ nearly zero. Therefore, for the sake of simplicity, we represent the vegetation with rigid stems in the model. As a results, $u_p$ is set to zero, and thus, $u_r = u$. We acknowledge that this limitation makes the somewhat species-specific and less universally applicable for other flexible vegetation (e.g., seagrass). Nevertheless, the methodology presented in this work can be applied to flexible vegetation by modeling the swaying motion of flexible stems and replacing the empirical equation used in this study, described later in Equation (5), with a formula that is developed for flexible vegetation (Anderson & Smith, 2014; Kobayashi et al., 1993; Lei & Nepf, 2019).

As discussed earlier, the vegetation drag coefficient is a function of flow and vegetation characteristics (Gijón Mancheño et al., 2021). In this study we adopt the T&N formula (Tanino & Nepf, 2008), shown below, and incorporate it into the XBNH model to approximate the vegetation drag coefficient, accounting for both flow dynamics and characteristics of the vegetation (Tanino and Nepf, 2008):

$$C_D = 2\left(\frac{\alpha_0}{R_p} + \alpha_1\right) \tag{5}$$

In this equation, $R_p$ stands for the plant Reynolds number, which is calculated as $u_v b_v / \nu$ where $u_v$ acting flow velocity on the submerged vegetation. Both $\alpha_0$ and $\alpha_1$ are empirical coefficients influenced by the volumetric fraction of the vegetation, denoted as $\varphi$. Utilizing empirical data from multiple studies (Koch & Ladd, 1997; Marsooli et al., 2016; Petryk, 1969; Tanino & Nepf, 2008), we applied a linear relationship between $\alpha_1$ and $\varphi$, expressed as $\alpha_1$ = 0.56 + 4.08 $\varphi$ with the $R^2$ value of 0.94. We also derived an analogous relationship between $\alpha_0$ and $\varphi$, formulated as $\alpha_0$= 5.26 + 318.1 $\varphi$, displaying an $R^2$value of 0.83 (Figure 2).

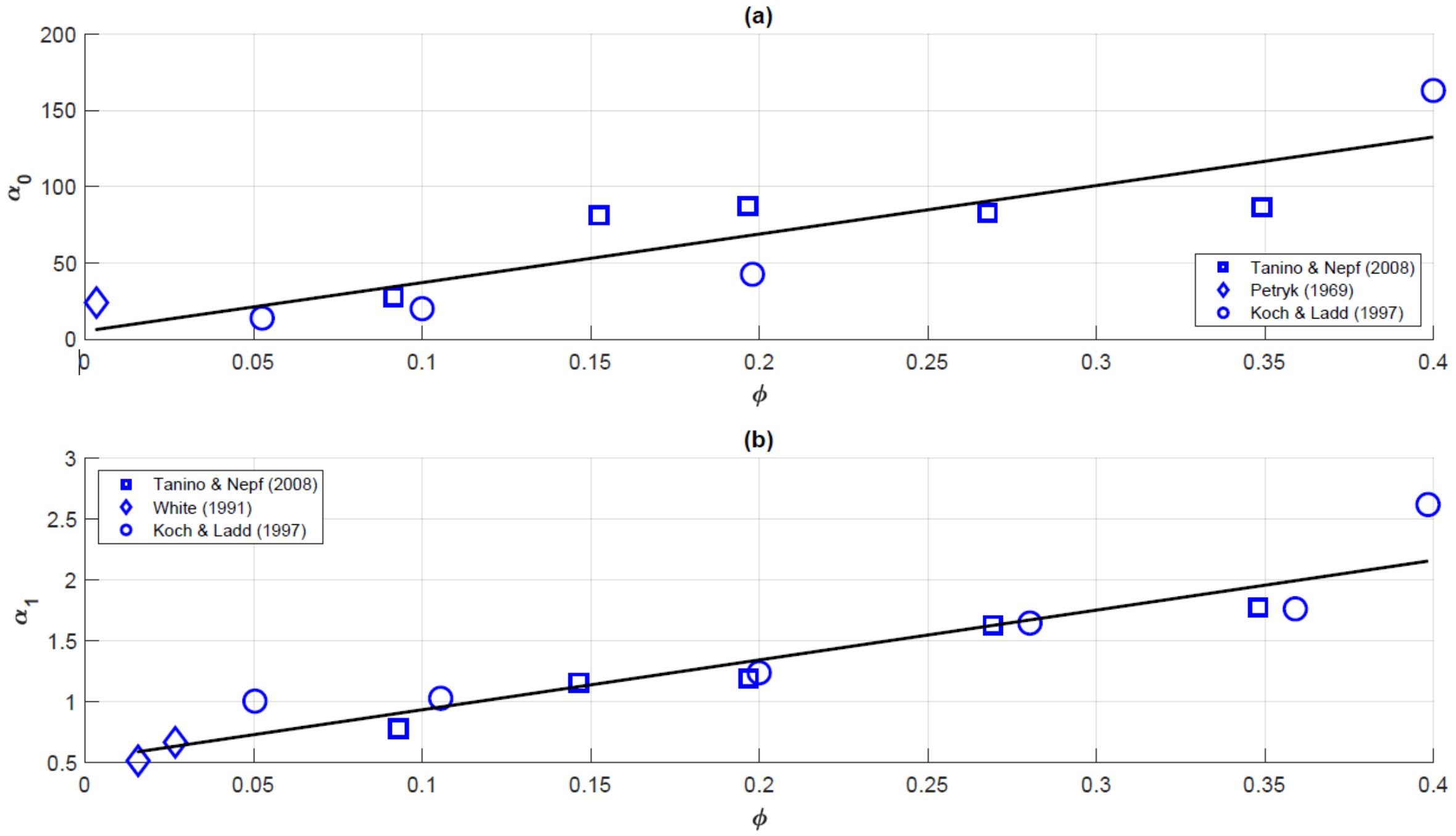

Figure 2. Scatter plot of empirical coefficients in the bulk drag coefficient formula of T&N, (a): $\alpha_0$, and (b): $\alpha_1$ over solid volume fraction ($\varphi$), based on the data collected from the literature (lines are linear regression model fitted to the data).

Calculating the plant Reynolds number $R_p$ for submerged vegetation requires corrections to the depth-averaged flow velocity obtained from the model. Unlike emergent vegetation, where the vegetation covers the entire water depth and therefore the depth-averaged flow velocity generally suffices for calculating $R_p$, submerged vegetation only covers a portion of the flow depth and consequently the depth-averaged velocity does not represent the flow velocity acting on the vegetation. The acting flow velocity $u_v$ on the submerged vegetation should be the velocity averaged over the height of the submerged vegetation layer (Figure 3). Previous research, such as that conducted by (W. Wu & Marsooli, 2012), has used a formulation initially derived by (Stone & Shen, 2002) analytical relations to determine this averaged velocity acting on the submerged vegetation. Their formulation proposes that $u_v$ can be calculated as $u\frac{\sqrt{h_v}}{d}$, where $h_v$ is the height of the submerged vegetation, as illustrated in Figure 3. We used this approach to adjust the velocity used in the plant Reynolds number calculation in the T&F formula.

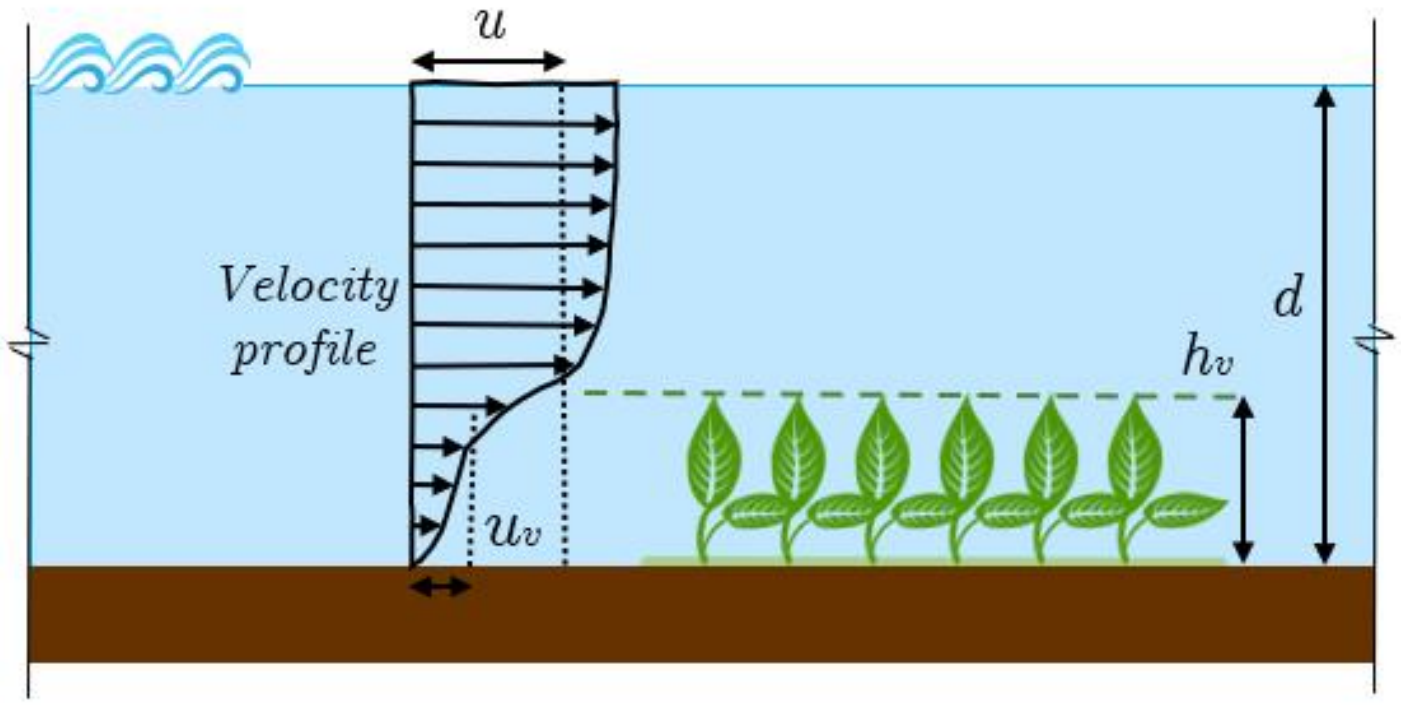

Figure 3. Assumption of (quasi-) linear velocity, as a superposition of individual mechanisms when flow passes the vegetation stems in submerged condition. *u* is the velocity averaged over the entire water depth. $u_v$ is the velocity averaged over the height of the vegetation.

The accuracy of XBNH model for modeling waves also relies on properly calibrating a series of input parameters associated with wave calculations. One such parameter is *maxbrsteep*, which sets a threshold for the rate of change in water surface elevation at wave breaking points, as discussed in (de Beer et al., 2021). Another calibrating parameter is *breakviscfac*, which modifies the horizontal viscosity caludalted by the Smagorinsky formulation (Smagorinsky, 1963). For this study, *maxbrsteep* was set to 0.65 and *breakviscfac* was set to 1.5 based on previous sensitivity analysis and validation of the XBNH model (Amini & Marsooli, 2023). The XBNH model applied here has undergone rigorous prior validation using small and large-scale laboratory experiments as well as field data, as described in Amini and Marsooli (2023). This provides confidence in the model's accuracy for simulating waves.

## 2.2. Laboratory Case Study

To analyze the effects of rigid vegetation on wave attenuation, we utilize laboratory measurements from (W. M. Wu et al., 2011). The laboratory experiments utilized a wave flume to test the effects of rigid model vegetation on wave height, breaking, and setup over a 1:21 sloping beach (Figure 4). The rigid model vegetation consisted of uniform cylindrical birch dowels with a diameter of 9.5 mm inserted vertically through holes in PVC sheets in a staggered arrangement. Wave gauge data was collected at multiple locations to quantify wave height transformation and mean water level profiles. Different irregular wave conditions were generated with a JONSWAP spectrum with significant wave heights of 3.7 cm to 7.9 cm and peak periods of 1.2 s to 1.8 s propagating over vegetated beach profile at a constant water depth of 40 cm. The details of the selected scenarios are presented in Table 1. In their experiments, scaled rigid vegetation with a density of $N_v$=3182 stems/m$^2$, stem diameter of $b_v$=3.2 mm, and height of $h_v$=20 cm were installed along the sloping beach profile with the spacing of λ=19.1 cm to represent a full-scale density of 350 stems/m$^2$ (Figure 5).

Table 1. Details of the irregular wave scenarios propagating over vegetated beach profile. $T_p$ is peak wave period, $H_i$ is incident wave height, and $H_s$ is significant wave height.

| Case number | $T_p$ (s) | $H_i$ (cm) | $H_s$ (cm) |
|---|---|---|---|
| I | 1.2 | 2.6 | 3.7 |
| II | 1.6 | 3.4 | 4.7 |
| III | 1.2 | 3.7 | 5.4 |
| IV | 1.8 | 5.7 | 7.9 |

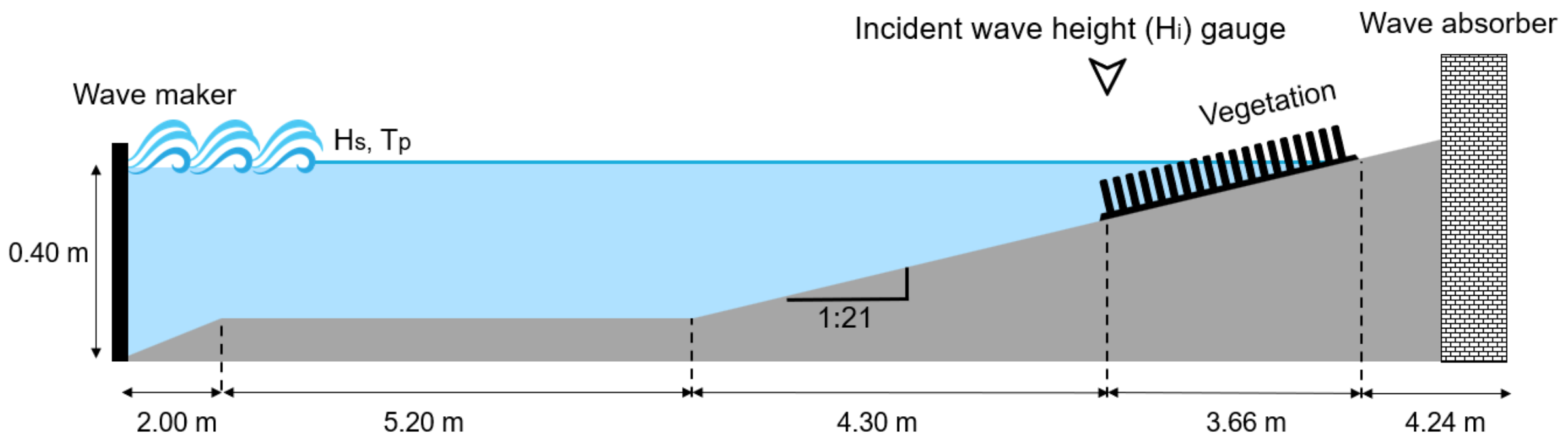

Figure 4. Wave flume and vegetation configuration of Wu experiments (W. M. Wu et al., 2011).

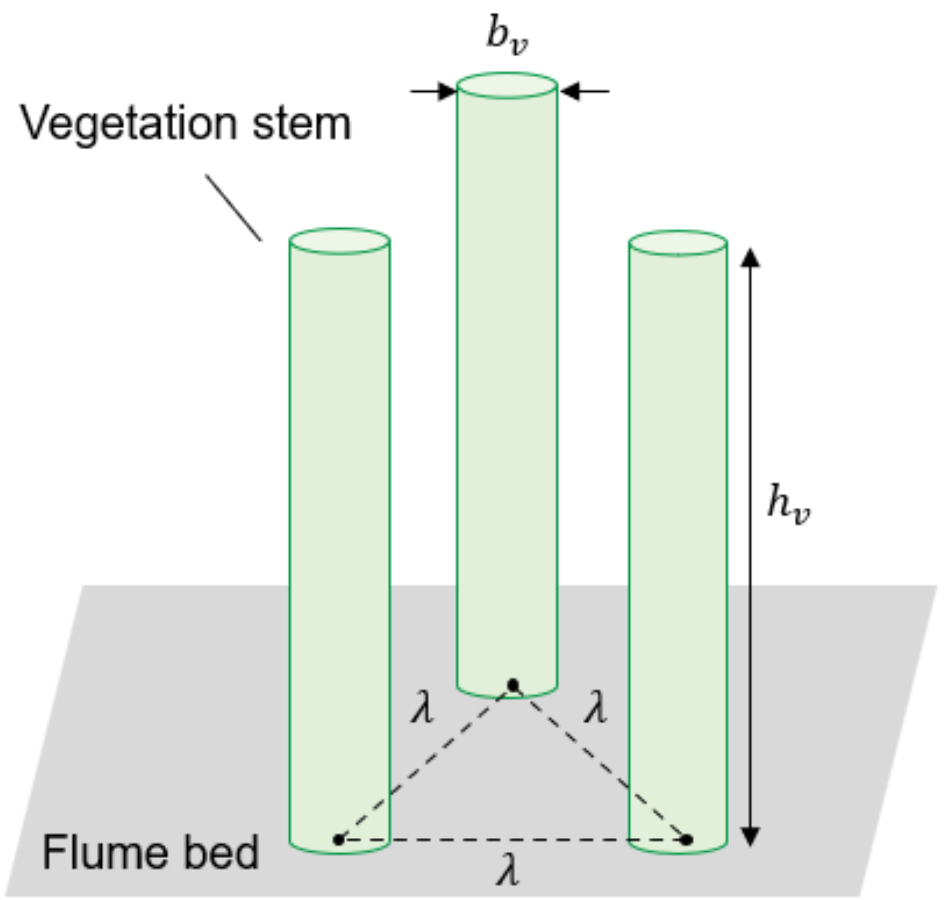

Figure 5. Schematic view of vegetation staggered arrangement utilized in the Wu experiments (W. Wu, 2012), where $\lambda$ is the center-to-center distance between individual stems, $h_v$ is stem height, and $b_v$ is the stem diameter.

## 2.3. Metaheuristic Optimization

Metaheuristic algorithms provide an efficient means to search for a large solution space and identify near-optimal solutions for problems with multiple variables, nonlinear constraints, and complex objective functions (Agushaka et al., 2023; Amini, Nasiri, et al., 2023; Johnvictor et al., 2022; Mirjalili, 2019; Wasim et al., 2022). In the present study, we incorporate two fast metaheuristic algorithms - Grey

Wolf Optimizer (GWO) (Mirjalili et al., 2014) and Moth-Flame Optimization (MFO) (Mirjalili, 2015) - to calibrate the vegetation drag coefficient ($C_d$) based on the laboratory measurements described in the previous sub-section.

GWO is a nature-inspired optimization algorithm based on simulating the hunting behavior and hierarchical social structure of grey wolf packs (Mirjalili et al., 2014). The algorithm models the leadership hierarchy within a wolf pack, with alpha, beta, delta, and omega wolves guiding the movements of the pack toward promising regions of the search space. This is mathematically represented by adjusting the positions of each search agent based on the fittest search agents at each iteration. The core strengths of GWO include rapid convergence, reduced computational cost, and avoidance of local optima traps (Makhadmeh et al., 2023; Negi et al., 2021).

The other approach utilized here, MFO, mimics the navigational strategy of moths using transverse orientation and spiraling movements to locate light sources, or flames (Mirjalili, 2015). The algorithm models moths as search agents navigating toward promising regions, with some degree of random adjustments to avoid entrapment in local optima. The spiral component enables wide exploration of the search space, while transverse orientation exploits promising areas. Key advantages of MFO are its simple implementation, fast convergence, and robustness (Shehab et al., 2020).

GWO and MFO have gained attention due to their unique characteristics and considerable performance in various real engineering problem domains compared with other popular optimization techniques such as Genetic Algorithms (GA) or Particle Swarm Optimization (PSO)(Dalmaz et al., 2023; Zhou et al., 2022). These methods are selected for two primary reasons. First, they aid achieving a harmonious equilibrium between exploration and exploitation (Al-Tashi et al., 2022). Take for example, the Moth-Flame Optimization (MFO), which adeptly combines random exploration and local exploitation, enabling it to effectively navigate the quest for optimal solutions. In contrast, widely used optimization techniques like Genetic Algorithms (GA) may encounter challenges in striking an intelligent balance. The selection process within GA, such as tournament or fitness proportionate selection, can impose a bias towards exploiting the highly fit individuals, consequently shrinking the exploration of the search space. As a result, the algorithm tends to concentrate on exploiting the more superior solutions, thereby limiting its exploration potential. Second, GWO and MFO allow incorporating both global and local search mechanisms. While crossover and mutation operators in GA contribute to global search, their impact on the population's diversity can vary. If the crossover rate is too high or the mutation rate is too low, the algorithm may prioritize local search and converge prematurely without thoroughly exploring the search space.

For this study, we develop an objective function that calculates the wave height attenuation over the vegetation, using a given $C_d$ value as input to the XBNH model. The objective function then computes the root-mean-square error (RMSE) between the modeled and measured wave heights over the vegetation patch. The goal is to minimize this RMSE by modifying $C_d$, thereby identifying the optimal $C_d$ that best aligns the modeled wave heights with the laboratory measurements.

$$\mathrm{RMSE} = \sqrt{\frac{\sum_{i=1}^{N} \left(H_i - \widehat{H}_i\right)^2}{N}} \tag{6}$$

Where N is number of measurements, $H_i$ is the modeled wave height over the vegetation patch, and $\widehat{H}_i$ is the measured wave height. The implementation of the GWO and MFO algorithms for $C_d$ optimization can be summarized in the following steps (Figure 6). Details of algorithms' evaluation processes are also provided by pseudo codes in the Appendix.

(i). Initialize the grey wolf pack (for GWO) or moth population (for MFO) with random $C_d$ values within a specified range.

(ii). Evaluate each search agent's $C_d$ value by running the XBNH model with that $C_d$ input and computing the RMSE between modeled and measured wave heights.

(iii). Update the positions of search agents based on GWO or MFO movement equations to move toward more optimal regions of the $C_d$ solution space.

(iv). Re-evaluate updated $C_d$ values via the objective function and update personal and global best solutions.

(v). Repeat steps iii-iv for a defined number of iterations or until convergence criteria are met.

(vi). Select the $C_d$ value associated with the global best solution as the optimized drag coefficient.

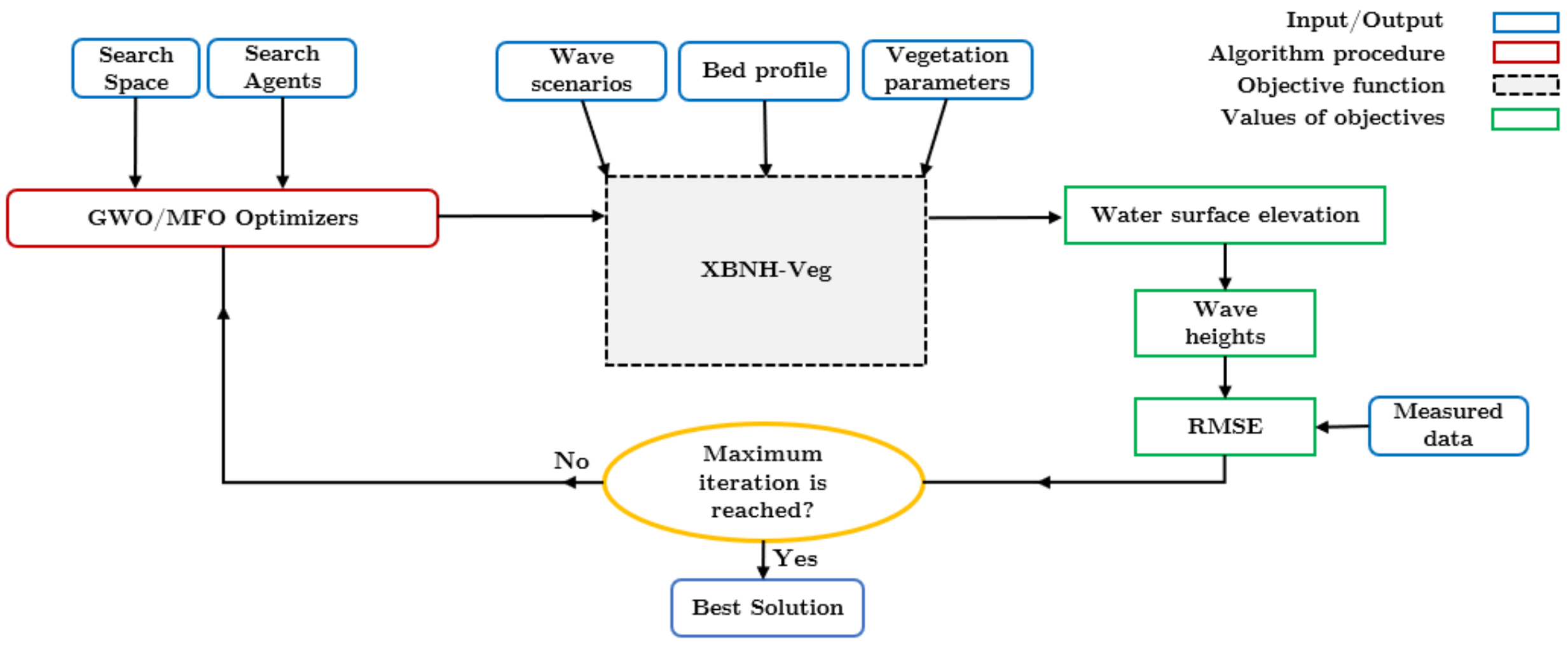

Figure 6. Flowchart of the drag coefficient ($C_d$) calibration utilizing grey wolf optimizer (GWO) and moth flame optimization (MFO) algorithm.

To determine the upper and lower limits of the search process, we use the drag coefficient estimation from the laboratory experiments of (W. Wu, 2012). For rigid model vegetation, under irregular waves, the estimated $C_d$ from measurements exhibited a decreasing trend as the Reynolds (Re) number increased (Figure 7). At higher Re values, the $C_d$ stabilized, approaching a constant minimum. In the measurements, no clear correlation was observed between the $C_d$ and the relative submergence depth ratio of plant height to water depth ($h_v/d$). The values of $C_d$ range from 1.8 to 10, and thus, we first selected this range as the search space in the optimization process, as shown in the highlighted band in the Figure 7. Less iteration would be needed to reach the satisfactory convergence by narrowing the search space. Our preliminary analysis showed that a wider search space makes more available room for search agents to explore outliers and given the local optima in very low and high values of $C_d$, they may result in an unfeasible answer. Thus, the constraint for the decision variable is considered as $1.8 \leq C_d \leq 10$.

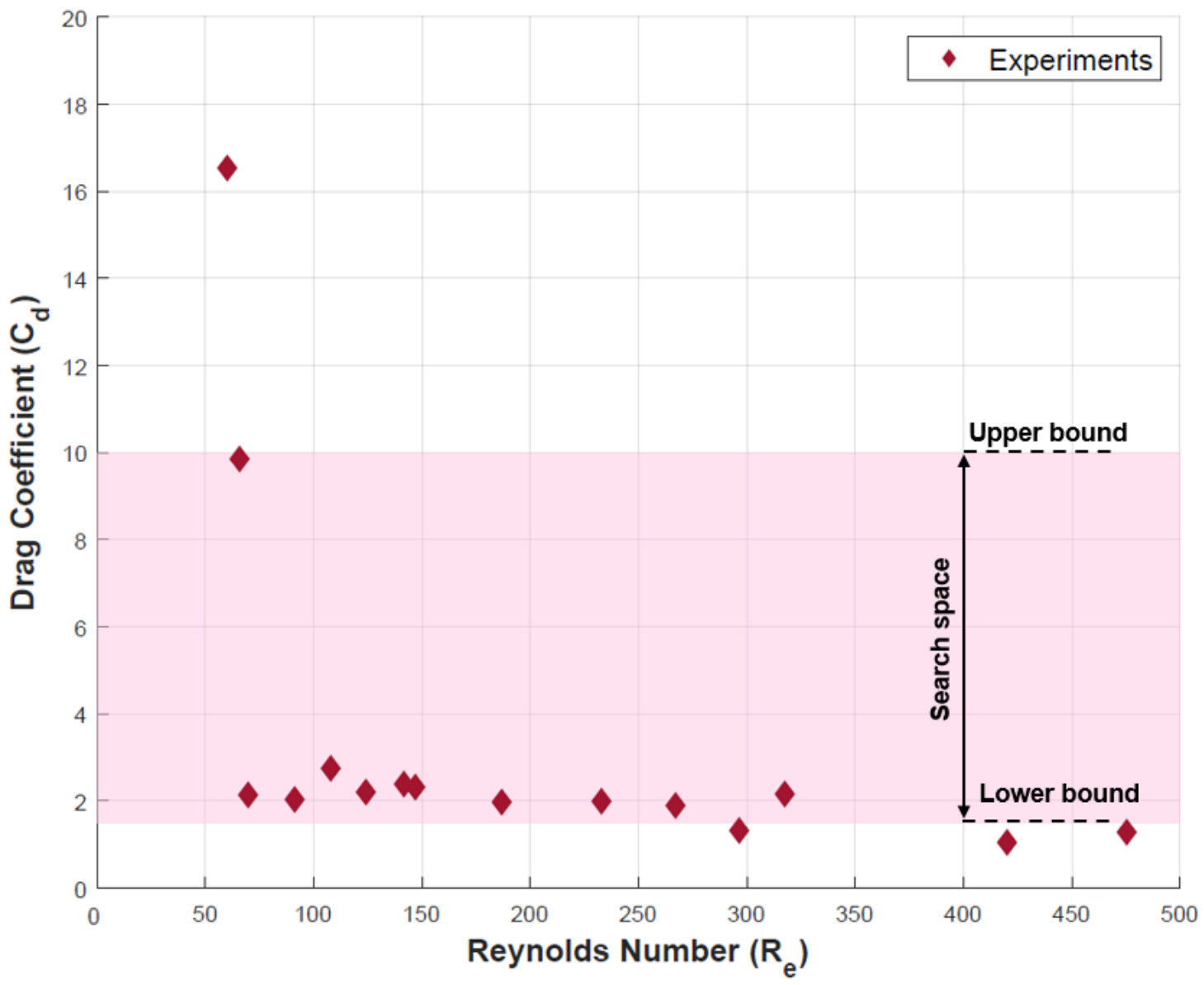

Figure 7. Drag coefficient from laboratory experiments of (W. Wu, 2012). 80% of the $C_d$ data fall within the highlighted range (1.8-10).

The key advantage of using metaheuristic algorithms like GWO and MFO is their ability to efficiently search the multi-modal solution space while avoiding the local optima (Al-Tashi et al., 2022; Sahoo & Saha, 2022). In our application, they automatically tune the $C_d$ coefficient for optimal wave height attenuation alignment. This overcomes the intensive effort and potential modeler biases involved in manual $C_d$ calibration. Additionally, the algorithms are inherently parallelizable, enabling calibration using high performance computing. By comparing the performance of GWO and MFO, we can also discern the most effective algorithm for this coastal engineering application.

## 3. Results and Discussion

This section presents and discusses the results obtained from each calibration approach. First, the manually calibrated drag coefficients are compared with those derived from the optimization algorithms. Following this, the performance of the Grey Wolf Optimizer and Moth-Flame Optimizer algorithms are evaluated and contrasted. Finally, the accuracy of all three methodologies in simulating wave height attenuation over vegetation is assessed. By comparing these distinct calibration techniques, we aim to determine the most effective strategy for calibrating the drag coefficient and modeling wave attenuation through coastal vegetation and show how the vegetation drag coefficient calibration processes can benefit from each of the techniques.

Figure 8 shows the normalized root-mean-square-error (RMSE) for manual calibration and optimization across the four test cases. For each case, we manually tested different drag coefficient ($C_d$) values between 1.8 and 2.8. The water surface elevation was obtained from the XBNH model, then wave heights were calculated using a zero-up crossing method. The wave heights from the numerical model and laboratory measurements were compared to calculate the RMSE. $C_d$ values above 2.8 are not simulated as the errors increased for all cases. The manual assessment also revealed characteristics of the search space landscape analysis for the optimization process. For instance, the feasible search range appeared unimodal overall, with a single global minimum. However, based on the convergence curves provided later, some local optima likely exist around the global minimum.

The MFO optimization algorithm achieved lower RMSE values in Cases I, II, while GWO performed slightly better in Case III and IV. This indicates MFO may be better suited for this application because it has faster convergence and more exploration capabilities compared to GWO. MFO has faster convergence compared to GWO due to its spiral mechanism to balance exploration and exploitation. The spiral component enables efficient traversal of the search space. MFO also has better capability to avoid getting stuck in local optima due to the randomization incorporated in its transverse orientation mechanism. Additionally, MFO requires fewer parameters to tune compared to GWO, making it simpler to implement and apply. For this application, the faster convergence, increased exploration, and simplicity of implementation make MFO more effective than GWO for calibrating the drag coefficient. Comparing optimization and manual calibration shows the optimization rapidly reaches satisfactory results given the search space complexity allows for efficient convergence due to the relatively smooth unimodal landscape. This could enable automated $C_d$ calibration without human expertise and avoid the intensive labor and potential inconsistencies of manual calibration.

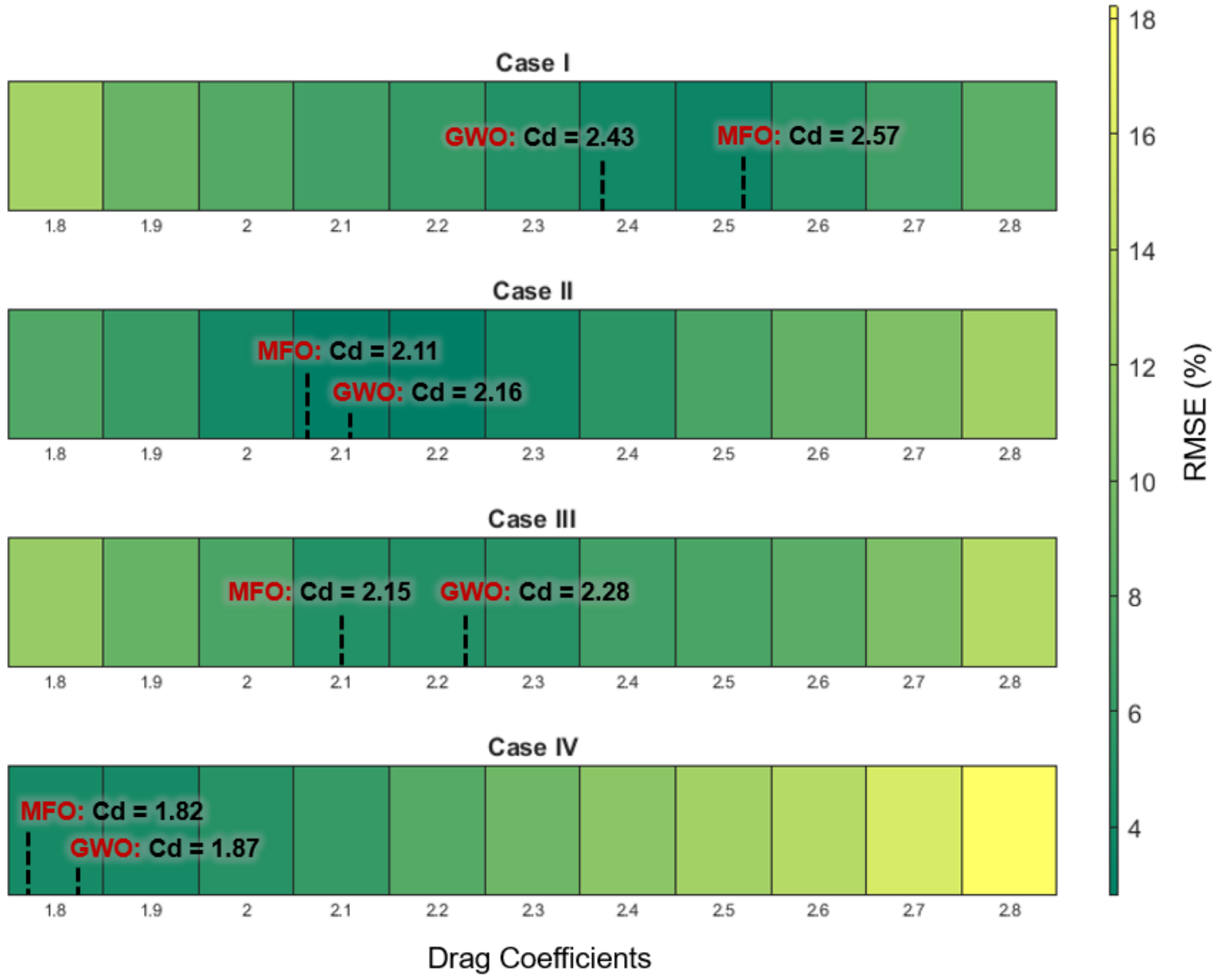

Figure 8. Landscape analysis of drag coefficient ($C_d$) using metaheuristic algorithms in four cases of the Wu experiments.

The exploration and performance of each optimization algorithm are visually represented in Figure 9. For this search, we deployed 10 search agents for each algorithm, with a cap set at 400 iterations. The figure's left panels show the convergence curve for each algorithm across different scenarios. Notably, the MFO exhibits rapid convergence early in the search across all cases, during the exploration phase. However, as the search progresses, GWO slightly overtakes the MFO in the exploitation phase in cases III and IV. This could be attributed to the inherent design of the GWO algorithm, which emphasizes late-stage optimization and precision, while MFO's strengths lie in early-stage broad search capabilities. A reassuring observation from our analysis was the satisfactory convergence of algorithms, showing an RMSE of less than 4.5% across all cases tested here. This indicated that the 400 iterations allocated were sufficient for our purpose. Any additional iterations would only amplify the computational burden without contributing significantly to the accuracy of the outcome.

The middle panels in Figure 9 offer a closer look at the GWO algorithm's strategy in pinpointing the optimal drag coefficient value. Initially, the algorithm casts a wide net, exploring the search space. However, upon discerning that increasing the $C_d$ doesn't necessarily lead to decreased RMSE values, the algorithm pivots to focus more on refining the optimal values it has already unearthed. This is evident, for instance, in case I, where the $C_d$ doesn't exceed 2.8. The exploration phase appears to plateau around the 200$^{th}$ iteration for all cases, as indicated by a reduction in the $C_d$ value fluctuations, especially pronounced in cases I and II. Following this, the algorithm transitions to its exploitation phase. A parallel trend is observable with the MFO algorithm (in the right panels Figure 9), although it transitions to the exploitation phase a tad earlier in cases I and II, a testament to its swifter convergence rate in these scenarios. Drawing from these observations, one could conclude that the MFO might be the superior choice, particularly in cases characterized by smaller waves (as in cases I and II). Further research could delve into the inherent algorithmic structures and parameters of MFO and GWO to glean deeper insights into their suitability across diverse real-world environmental conditions including wave conditions observed in the field.

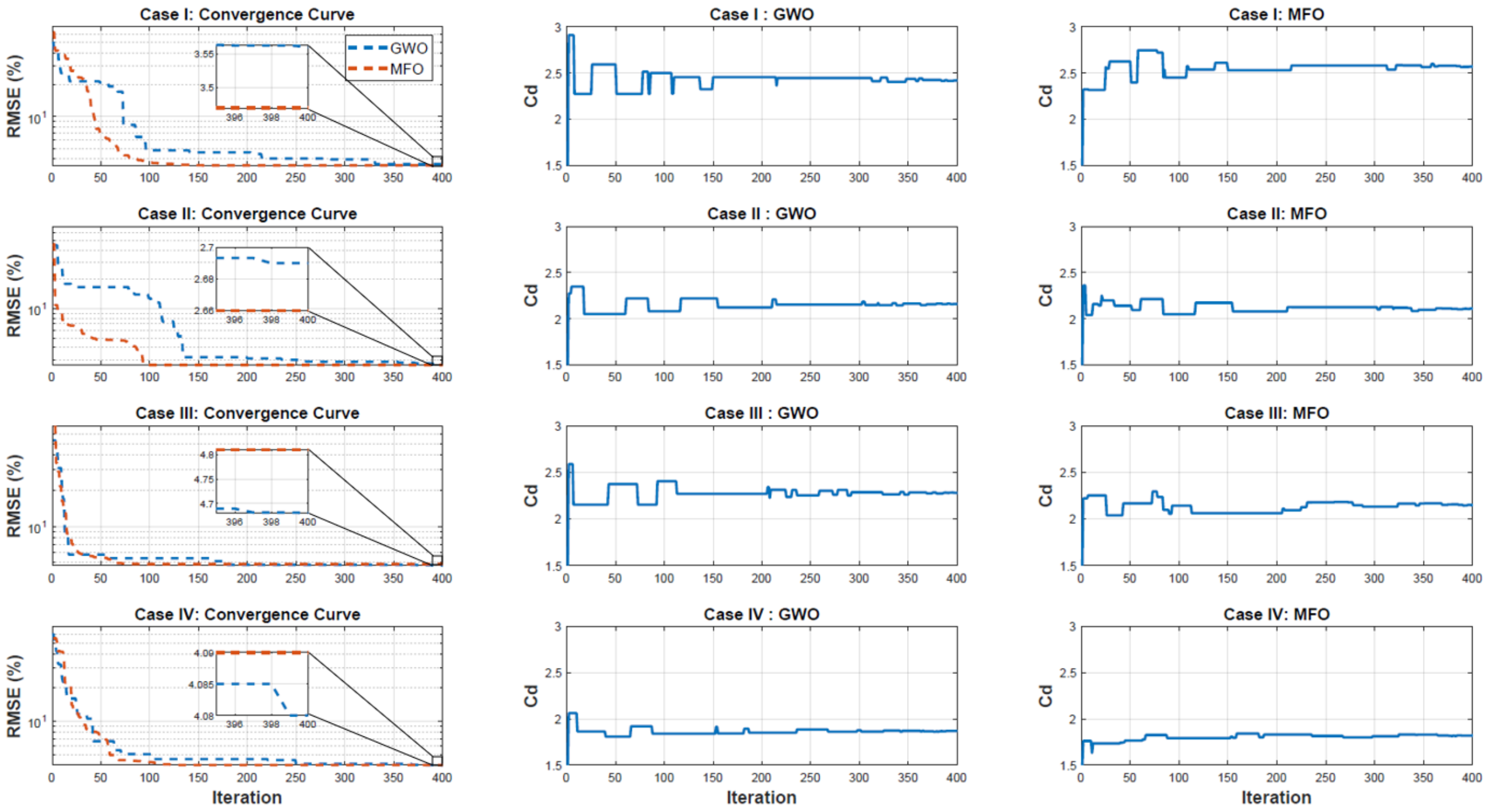

Figure 9. Left panels: convergence curve of metaheuristic methods over 400 iterations trying to minimize the RMSE. Middle and right panels: searching process of GWO and MFO algorithms spanning over the drag coefficient range.

Figure 10 shows an in-depth comparison of all three calibration approaches explored in this study for evaluating wave height attenuation as waves pass through the vegetation. Table 2 summarizes the RMSE of results obtained from each calibration approach. While the T&N bulk drag coefficient formula is appealing due to its ease of implementation and its dynamic calculation of $C_d$ – which accounts for the variation in environmental conditions – its accuracy is slightly wanting across all cases. Nevertheless, this minor discrepancy can be overlooked, given that this approach doesn't mandate measurements for calibrating the vegetation drag coefficient. Notably, the disparity in accuracy of T&N formula is much smaller in Case IV, characterized by a larger wave with a longer period. The figure reinforces the observation that the T&N formula's accuracy is commendable within the initial segment of the vegetation patch. However, as the waves pass through the vegetation, the accuracy reduces. This can be due to the T&N formula being developed and validated for a limited range of plant Reynolds numbers between 40-685, as noted in the original study by Tanino and Nepf (2008). The conditions in Case IV likely stay within this Reynolds number range longer through the vegetation patch compared to the other smaller wave cases, leading to the smaller disparity in accuracy. Nevertheless, the results derived from the T&N approach still remain within a small RMSE range, as detailed in the table for all scenarios.

Table 2. Root-mean-square-error (RMSE) of the different approaches across four cases.

| Case number | Manual calibration | T&F vegetation drag formula | GWO | MFO |
|---|---|---|---|---|
| **I** | 3.49 % | 5.91 % | 3.56 % | 3.47 % |
| **II** | 2.81 % | 2.86 % | 2.69 % | 2.66 % |
| **III** | 4.87 % | 5.45 % | 4.68 % | 4.81 % |
| **IV** | 4.17 % | 4.16 % | 4.08 % | 4.09 % |

Building upon these findings, it's pivotal to recognize the inherent trade-offs in choosing one approach over another. The primary advantage of an empirical formula, e.g., the T&N formula, lies in its adaptability to hydrodynamic changes. Furthermore, the formula can be used to estimate the drag coefficient without the need for measurements. This is particularly invaluable in real-world applications where data acquisition may pose challenges. On the other hand, the T&N formula adopted in this study was developed and validated based on laboratory experiments with rigid cylinder arrays representing emergent vegetation over a limited range of Reynolds numbers. Therefore, applying the formula outside of these conditions at much higher or lower Reynolds numbers may require adjustments to the empirical coefficients to ensure accuracy. The formula's applicability is also constrained by the cylinder array setup used in its derivation and may need modifications when applied to other canopy structures. Overall, while an empirical formula can provide a convenient and computationally inexpensive approach to estimate the vegetation drag coefficient, its limited range of environmental conditions should be considered when applying it more broadly. However, the applicability and adaptive nature of an empirical formula within the acceptable range, combined with its reasonable accuracy, positions it as a potential choice for estimating the vegetation drag coefficient.

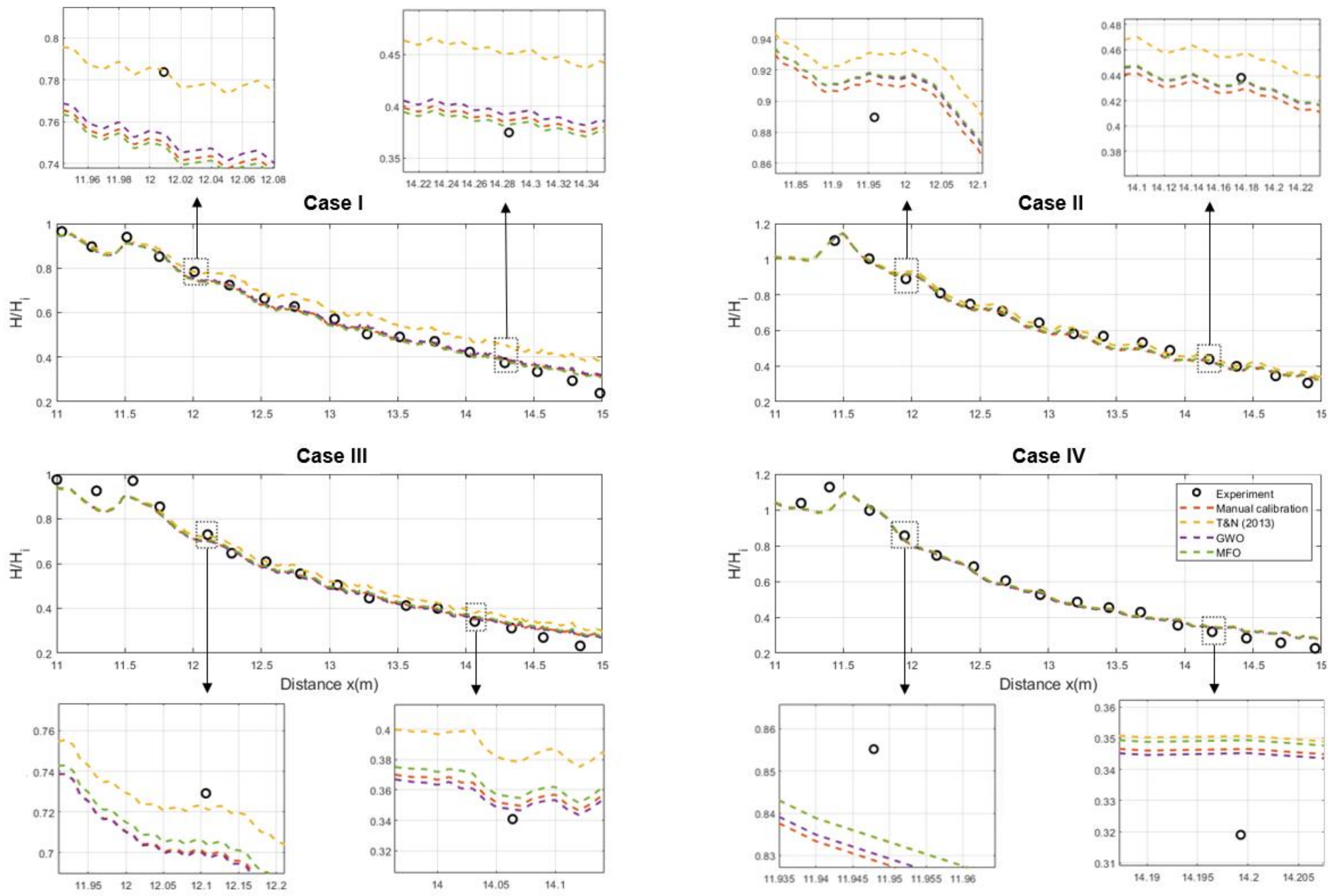

Figure 10. Comparison of dimensionless wave height attenuation due to the vegetation represented by $C_d$ values calculated manually, by T&F formula, GWO, and MFO optimization. The experimental values for non-dimensional wave heights, depicted as scatters. The line plots show the outcomes derived from the manual, empirical, and optimal calibration of $C_d$. Wave heights are normalized by the incident wave height, $H_i$.

## 4. Conclusion

This research evaluated three distinct methodologies for estimating the vegetation drag coefficient ($C_d$) in simulations of wave height attenuation by nearshore rigid vegetation. The methodologies included traditional manual calibration, integration of metaheuristic optimization algorithms with hydrodynamic modeling, and an empirical bulk drag coefficient formulation (Tanino & Nepf, 2008). Each approach was tested against laboratory measurements of wave propagation over a sloping vegetated beach.

The key findings reveal the empirical bulk coefficient formula yields satisfactory performance once incorporated into the XBNH model. While marginally less accurate than optimization algorithms in some instances, it provided reasonable accuracy across all test cases, while it is adaptive to hydrodynamic conditions and easy to implement in numerical models. The optimization algorithms, Grey Wolf Optimizer (GWO) and Moth-Flame Optimizer (MFO), also performed well, rapidly converging to aligned $C_d$ values. MFO displayed faster convergence and broader search space exploration, making it better suited than GWO for this application. However, both algorithms may require high computational resources to run over high number of iterations. The empirical formula offers comparable accuracy than optimization, without the resource demands. Besides, manually calibrated coefficients aligned closely with those derived from optimization. However, manual calibration can be labor and time-intensive (as is the optimization method) and subject to human error. The integration of the bulk coefficient formula and optimization algorithms automate parts of the process.

The key innovation of this study is the novel incorporation of metaheuristic optimization algorithms, which enables efficient automated searches to identify optimal $C_d$ values aligned with experimental data. The Grey Wolf and Moth-Flame algorithms rapidly converged to precise $C_d$ coefficients, overcoming the intensive modeler labor and inconsistencies of manual calibration. This represents a major advance by harnessing optimization techniques to improve the accuracy, efficiency, and consistency of $C_d$ calibration. While the integration of the empirical formula also yielded good performance, the optimization algorithms exemplify the profound potential of computational methods to transform traditional manual calibration practices to an automated approach. Testing these algorithms provides a framework to guide methodology selection based on use case constraints. The algorithms' scalability and transferability offer ample prospects for advancing real-world coastal vegetation modeling across diverse scenarios. Further refinement of optimization methods and empirical formula adaptations hold promise for continued enhancements in future studies.

## 5. Appendix

This appendix provides the pseudo code for the two metaheuristic optimization algorithms used in this study - Grey Wolf Optimizer (GWO) and Moth-Flame Optimization (MFO). The GWO algorithm (Mirjalili et al., 2014) models the hunting and leadership behaviors of grey wolf packs. The position of each search agent represents a potential solution. The algorithm iteratively updates the agent positions based on the locations of the fittest search agents, representing the alpha, beta, and delta wolves. This enables the pack to converge on the optimal solution. The pseudo code is provided in Algorithm 1.

**Algorithm 1:** Gray Wolf Optimizer

```
Procedure GWO( ):
  N=10, Max_iter=400                  ▹ Population size and Maximum number of iterations
  𝕊 = {⟨Cd_1⟩, …⟨Cd_2⟩} ⇒ lb_1^N ≤ 𝕊 ≤ ub^N
Initialize parameters                 ▹ Alpha, Beta, Delta
Wolves_sorted = sort(Wolves)
for iter in [1, …, Max_iter] do
   for i in [1, …, N] do
      for j in [1, …, M] do
         A1 = 2 * rand() - 1
         C1 = 2 * rand()
         D_alpha = abs(C1 * Alpha(j) - Wolves(i, j))
         X1 = Alpha(j) - A1 * D_alpha
         A2 = 2 * rand() - 1
         C2 = 2 * rand()
         D_beta = abs(C2 * Beta(j) - Wolves(i, j))
         X2 = Beta(j) - A2 * D_beta
         A3 = 2 * rand() - 1
         C3 = 2 * rand()
          D_delta = abs(C3 * Delta(j) – Wolves(i, j))
          X3 = Delta(j) - A3 * D_delta
         Wolves(i, j) = (X1 + X2 + X3) / 3
      end for
   end for
   Evaluate the fitness of all wolves
   Update Alpha, Beta, and Delta based on the top three solutions
end for
EvaluateBest(Wolves)                  ▹ Compute the best fitness
return Wolves, best_positions
end Procedure
```

The MFO algorithm (Mirjalili, 2015) mimics the navigation strategy of moths using transverse orientation and spiral movements to locate light sources. Each agent's position denotes a candidate solution. The algorithm balances exploration and exploitation by updating agent locations using components of transverse orientation, spiral motion, and randomization. The pseudo code is shown in Algorithm 2.

**Algorithm 2:** Moth Flame Optimization

```
Procedure MFO( ):
  N=10, Max_iter=400                         ▹ Population size and Maximum number of iterations
  $\mathbb{S} = \{\langle Cd_1\rangle, \ldots \langle Cd_2\rangle\} \Rightarrow lb_1^N \le \mathbb{S} \le ub^N$
Initialize parameters
Moths_sorted = sort(Moths)
  for iter in [1, …, Max_iter] do
    b = 1                                    ▹ Constant
    t = iter / Max_iter  Time factor
    for i in [1, …, N] do
      for j in [1, …, M] do
        distance_to_flame = abs(Moths_sorted(i, j) - Moths(i, j))
        t_d = (a - 1) * t + 1
        Moths(i, j) = distance_to_flame * exp(b * t_d) * cos(t_d * 2 * π) + Moths_sorted(i, j)
      end for
    end for
    Evaluate the fitness of all moths
    Sort moths and flames based on fitness
    if N > Flame_no then
      Replace the worst moths with the best flames
    end if
  end for
EvaluateBest(Moths)                          ▹ Compute the best fitness
return Moths, best_positions
end Procedure
```

## 6. Acknowledgement

This work results from research sponsored by the Engineer Research and Development Center-Coastal and Hydraulics Laboratory (ERDC-CHL) under Contract No. W912HZ22C0010.

## 7. CRediT author statement

**Erfan Amini**: Conceptualization, Methodology, Software, Investigation, Validation, Writing - Original Draft, Visualization. **Reza Marsooli**: Conceptualization, Methodology, Resources, Writing - Review & Editing, Supervision, Funding acquisition. **Mehdi Neshat**: Writing - Review & Editing, Supervision.